\documentclass[aip,preprint,onecolumn]{revtex4-1}
\usepackage{graphicx}
\usepackage{dcolumn}% Align table columns on decimal point
\usepackage{bm}% bold math
\usepackage{color}

\begin{document}

% Use the \preprint command to place your local institutional report number
% on the title page in preprint mode.
% Multiple \preprint commands are allowed.
%\preprint{}

\title{Dynamics of viscous coalescing droplets in a saturated vapor phase} %Title of paper
%Viscous coalescence of liquid droplets in a saturated vapor phase
% repeat the \author .. \affiliation  etc. as needed
% \email, \thanks, \homepage, \altaffiliation all apply to the current author.
% Explanatory text should go in the []'s,
% actual e-mail address or url should go in the {}'s for \email and \homepage.
% Please use the appropriate macro for the type of information

% \affiliation command applies to all authors since the last \affiliation command.
% The \affiliation command should follow the other information.
\author{Lina Baroudi}
\affiliation{Department of Mechanical Engineering, City College of City University of New York, New York, NY 10031, USA}
\author{Sidney R. Nagel}
\affiliation{James Franck Institute and Department of Physics, University of Chicago, Chicago, IL 60637, USA}
\author{Jeffrey F. Morris}
\affiliation{Department of Chemical Engineering and Levich Institute, City College of City University of New York, New York, NY 10031, USA}
%\author{Sidney R. Nagel}
%\affiliation{James Franck Institute and Department of Physics, University of Chicago, Chicago, IL 60637, USA}
%\author{Jeffrey F. Morris}
%\affiliation{Department of Chemical Engineering and Levich Institute, City College of City University of New York, New York, NY 10031, USA}
\author{Taehun Lee}
\email{thlee@ccny.cuny.edu}
\affiliation{Department of Mechanical Engineering, City College of City University of New York, New York, NY 10031, USA}

%\email[]{Your e-mail address}
%\homepage[]{Your web page}
%\thanks{}
%\altaffiliation{}
% Collaboration name, if desired (requires use of superscriptaddress option in \documentclass).
% \noaffiliation is required (may also be used with the \author command).
%\collaboration{}
%\noaffiliation

%\date{\today}

\begin{abstract}
The dynamics of two liquid droplets coalescing in their saturated vapor phase are investigated by Lattice Boltzmann numerical simulations. Attention is paid to the effect of the vapor phase on the formation and growth dynamics of the liquid bridge in the viscous regime. We observe that the onset of the coalescence occurs earlier and the expansion of the bridge initially proceeds faster when the coalescence takes place in a saturated vapor compared to the coalescence in a non-condensable gas. We argue that the initially faster evolution of the coalescence in the saturated vapor is caused by the vapor transport through condensation during the early stages of the coalescence.

\end{abstract}

%\pacs{}% insert suggested PACS numbers in braces on next line

 %\maketitle must follow title, authors, abstract and \pacs
\let\newpage\relax\maketitle
%\maketitle
%\newcommand{\inlinemaketitle}{{\let\newpage\relax\maketitle}}
%\inlinemaketitle

% Body of paper goes here. Use proper sectioning commands.
% References should be done using the \cite, \ref, and \label commands
%\section{}
%\label{}
%\subsection{}
%\subsubsection{}

% If in two-column mode, this environment will change to single-column format so that long equations can be displayed.
% Use only when necessary.
%\begin{widetext}
%$$\mbox{put long equation here}$$
%\end{widetext}

\begin{figure}
\centering
\includegraphics[scale=0.4]{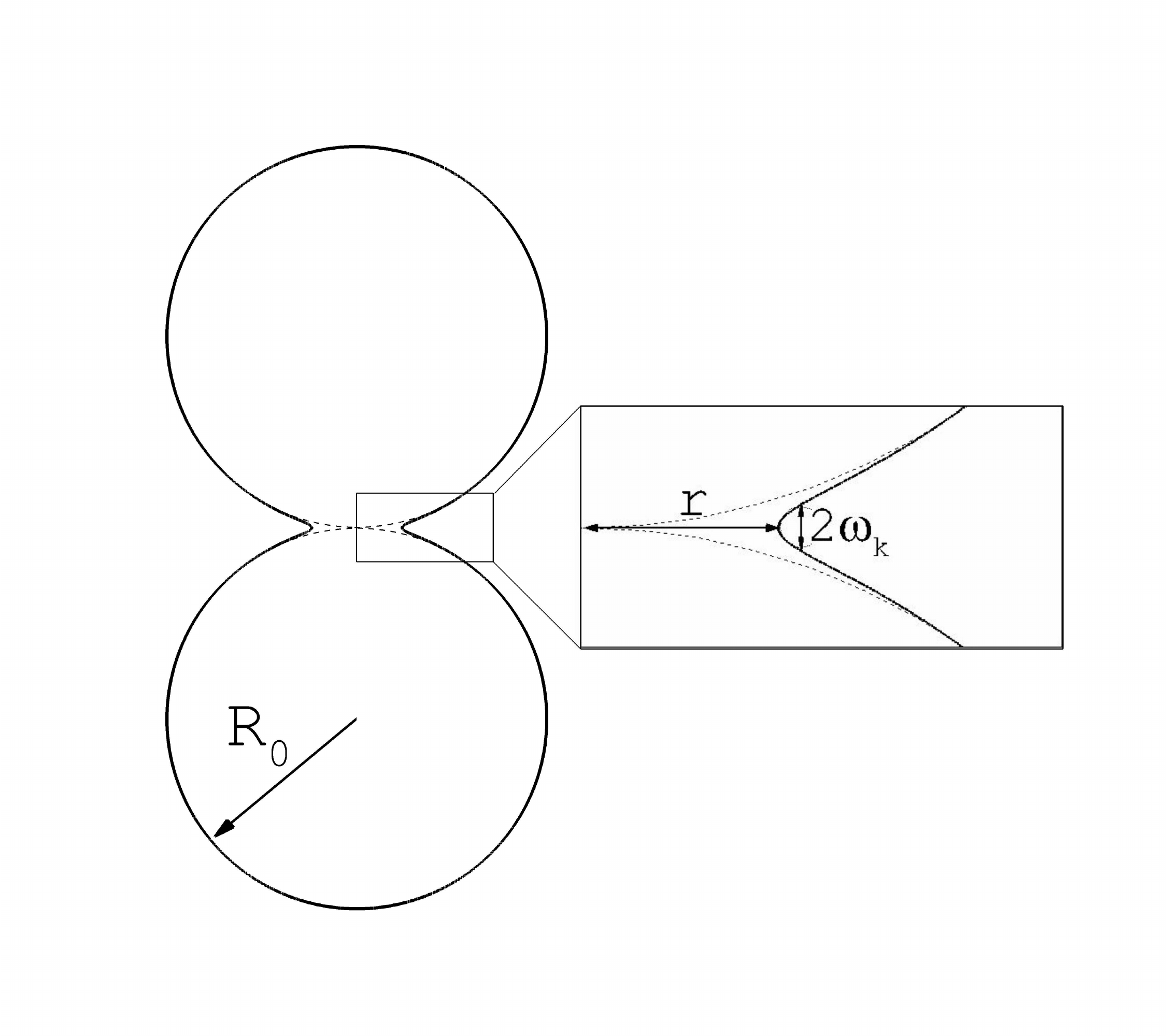}
\caption{Sketch of the coalescence of two identical drops. $R_0$ is the drop radius, $r$ is the bridge radius, and $\omega_{k}$ is the meniscus radius of curvature.}
\label{fig:sketch}
\end{figure}

Drop coalescence is of major importance in many industrial applications and environmental contexts~\cite{Ashgriz-1990,Hajra_2003, Hu-1995,Pruppacher_2010}. When two liquid drops come into contact, a microscopic liquid bridge forms between them and rapidly expands until the two drops merge into a single bigger drop. Numerous studies~\cite{Hopper_1984,Egger_1999,Duchemin_2003,Aarts_2005,Yao_2005,Lee_2006,Paulsen_2011,Paulsen_2012,Sprittles_2012,Baroudi_2014, Sprittles_2014} have been devoted to the investigation of the coalescence singularity in the case where the drops coalesce in vacuum or air. These studies aimed to understand the dynamics governing the outward motion of the liquid bridge radius $r(t)$ with time (Fig.~\ref{fig:sketch}). Considering the forces that govern the coalescence dynamics, the coalescence process has been classified into two regimes with a cross-over region between them: one is a viscous regime, where the viscous forces oppose the expansion of the liquid bridge formed between the drops, and the other is an inertial regime, where the bridge motion is opposed by inertial forces. Recently Paulsen \textit{et al.}~\cite{Paulsen_2012} argued that for all drop viscosities, the coalescence process starts in the ``inertially limited" viscous regime, in which inertial and viscous forces play a role in the dynamics. In all of the aforementioned studies the coalescence processes were taking place in a medium of negligible vapor pressure (non-condensable gas), and were assumed to occur at a constant liquid volume. However, coalescence of liquid drops may also take place in a medium of relatively high vapor pressure (condensable vapor phase), where the effect of the surrounding vapor phase should not be neglected, such as the merging of  drops in clouds~\cite{Pruppacher_2010} and in coalescing filters~\cite{Hajra_2003}. Here, we carry out Lattice Boltzmann numerical simulations to investigate the dynamics of viscous coalescence in a saturated vapor phase.
\par

In a non-condensable medium, the coalescence dynamics is determined by the dimensionless bridge radius $r(t)/R_0$ and Ohnesorge number, $Oh=\eta_{l}/\left(\rho_{l}\sigma R_0\right)^{1/2}=\left(r/R_0 Re\right)^{1/2}$, where $R_0$ is the drop radius, $\eta_{l}$ is the liquid viscosity, $\rho_{l}$ is the liquid density, $\sigma$ is the surface tension, and $Re$ is the Reynolds number for the flow near the liquid bridge, defined as $Re = \rho_{l}\sigma r/\eta_{l}^{2}$. For all simulations to be described here, $Re<1$ during the time of interest. The propagation of the liquid bridge with time can be described by a power law, allowing prediction of the time required for coalescence.  In the viscous regime, coalescence is driven by capillary forces and opposed by viscous forces, and a simple scaling argument will lead to a linear scaling of the bridge radius $r(t)/R_{0}\propto t/t_{v}$, where the time $t$ is measured from the moment of contact and $t_{v} \sim \eta_l R_0/\sigma$ is the viscous time scale.  The full theory~\cite{Egger_1999} predicts a logarithmic correction, $r(t)/R_{0}\propto (t/t_{v})ln(t/t_{v})$. Subsequent experiments have reported $r(t)/R_{0}\propto t/t_{v}$ in the viscous regime, with no evidence for the predicted logarithmic correction. Paulsen \textit{et al.}~\cite{Paulsen_2012} argued that the linearly expanding liquid bridge observed in previous experiments has been incorrectly assumed to represent viscous regime dynamics; they assert that this linear evolution represents the dynamics in the ``inertially limited'' viscous regime.\par

To investigate how the inclusion of the saturated vapor phase affects the viscous coalescence dynamics and the growth behavior of the liquid bridge, we study the coalescence of two resting liquid drops in a saturated vapor phase and their coalescence in a non-condensable gas, and compare the resulting dynamics. We consider two lattice Boltzmann equation (LBE) approaches to model the coalescence process.
The first is a non-ideal fluid LBE model~\cite{Lee_2006}, which is employed to model the coalescence of liquid drops in their saturated vapor phase. The second is a two phase fluid (nearly) incompressible LBE model~\cite{Lee2005}, and it is employed to model the coalescence of liquid drops in a non-condensable gas. The numerical results of the coalescence in a non-condensable gas will be compared to the available experimental results. The macroscopic continuity and momentum equations recovered from the LBE models through the Chapman-Enskog expansion are $\partial_t \rho + \nabla\cdot\left(\rho {\bf u}\right) = 0$ and $\partial_t \left(\rho {\bf u}\right)+\nabla\cdot\left(\rho {\bf uu}\right)=-\nabla p +\kappa\rho\nabla\nabla^2 \rho+\nabla\cdot{\bf \Pi} $ . Here $\rho$ is the density, $\bf u$ is the macroscopic velocity, $p$ is the pressure, $\kappa$ is the gradient parameter, and $\bf \Pi$ is the viscous stress tensor.
In essence, these two LBE models differ only by the way the pressure is updated. In both models, the pressure $p$ is obtained from the isothermal pressure evolution equation, which is given by $\partial_t p+\rho c_s^2\nabla\cdot{\bf u}+{\bf u}\cdot\nabla p = 0$, where $c_{s}^{2}=\left(\partial_\rho p\right)_s$ is the square of speed of sound at constant entropy (s). In the non-ideal fluid LBE model, $\partial_\rho p$ for a typical cubic equation of state (EOS) is not constant and turns negative at the phase interfaces, which may trigger isothermal phase change due to pressure variation~\cite{Lee2003}. In the incompressible LBE model, $\partial_\rho p$ is assumed to be constant and positive, and consequently, phase change is not allowed to take place. Detailed implementation of the models is described elsewhere~\cite{Lee_2006,Lee2005}.

In our numerical simulations, the coalescence of a pair of two-dimensional resting liquid drops of identical radii $R_0= 400$ lattice units (l.u.) generated on a $3000\times2000$ l.u. periodic computational domain is studied. Two and three-dimensional coalescence are expected to be equivalent to leading order~\cite{Egger_1999,Paulsen_2012,Sprittles_2014,Burton_2007}. Grid dependency and domain size dependency tests are performed using different grid resolutions and domain sizes. For the droplet radius $R_{0} = 400$ l.u. used in our simulations, the results are essentially independent of increased grid resolution, and for the domain of size $3000\times2000$ l.u. increased size of the domain has negligible influence on the evolution of the liquid bridge. The initial separation between the drops is $\delta=2$ l.u., and the Cahn number $Cn = \epsilon/2R_0 = 4/800= 0.005$, where $\epsilon$ is the interface thickness. Both $\delta$ and $Cn$ are fixed in all the simulations.

\begin{figure}
\centering
%\hspace{-2cm}
%incomp_before_touch_ZOOM_big_sep_no_color.eps
%before_touch_ZOOM_big_sep_div.eps
\begin{minipage}[t]{2.cm} %{0.2\linewidth}
\includegraphics[scale=0.28]{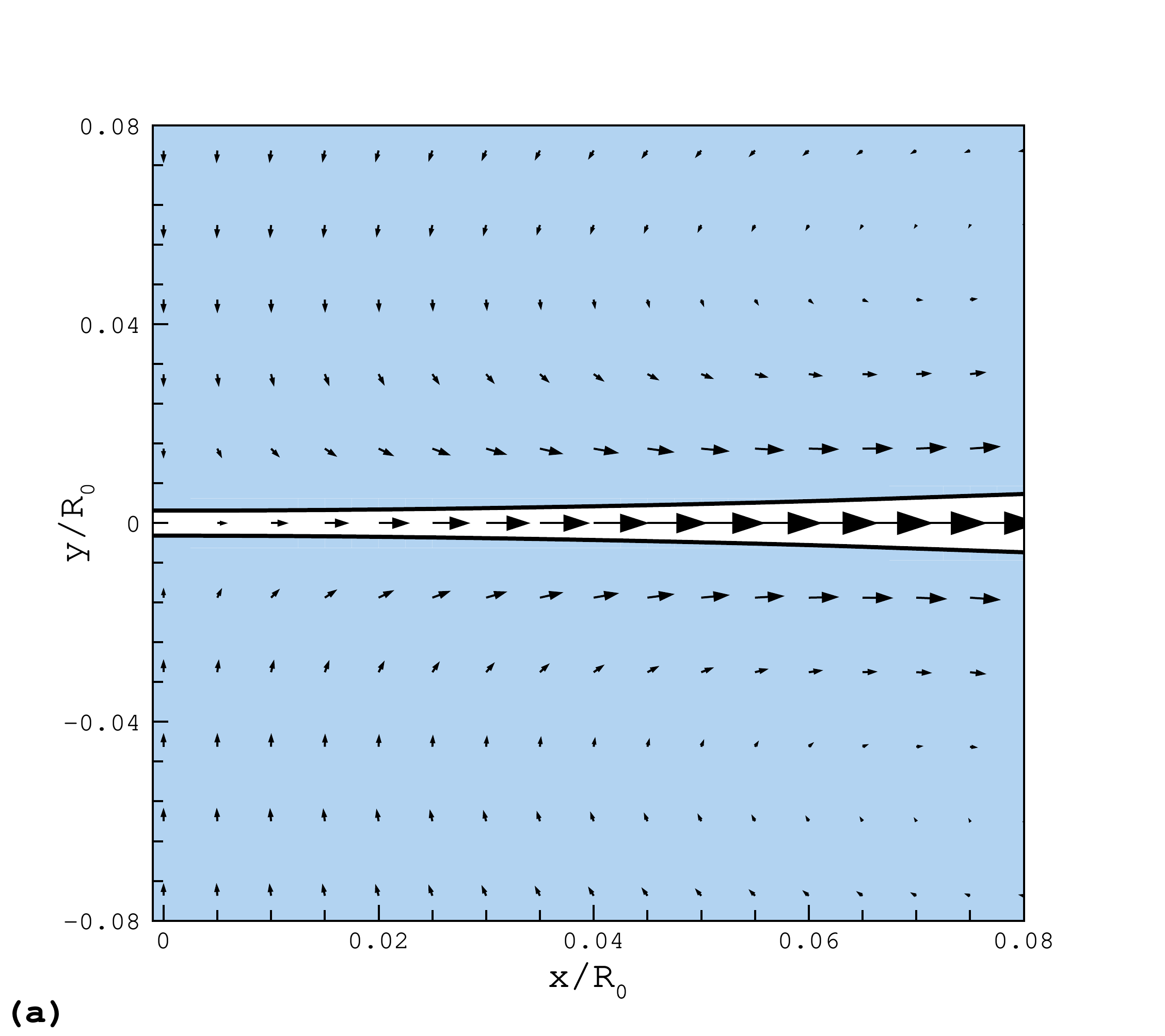}
\end{minipage}
\hspace{5cm}
\begin{minipage}[t]{5.cm}     %{0.2\linewidth}
\includegraphics[scale=0.28]{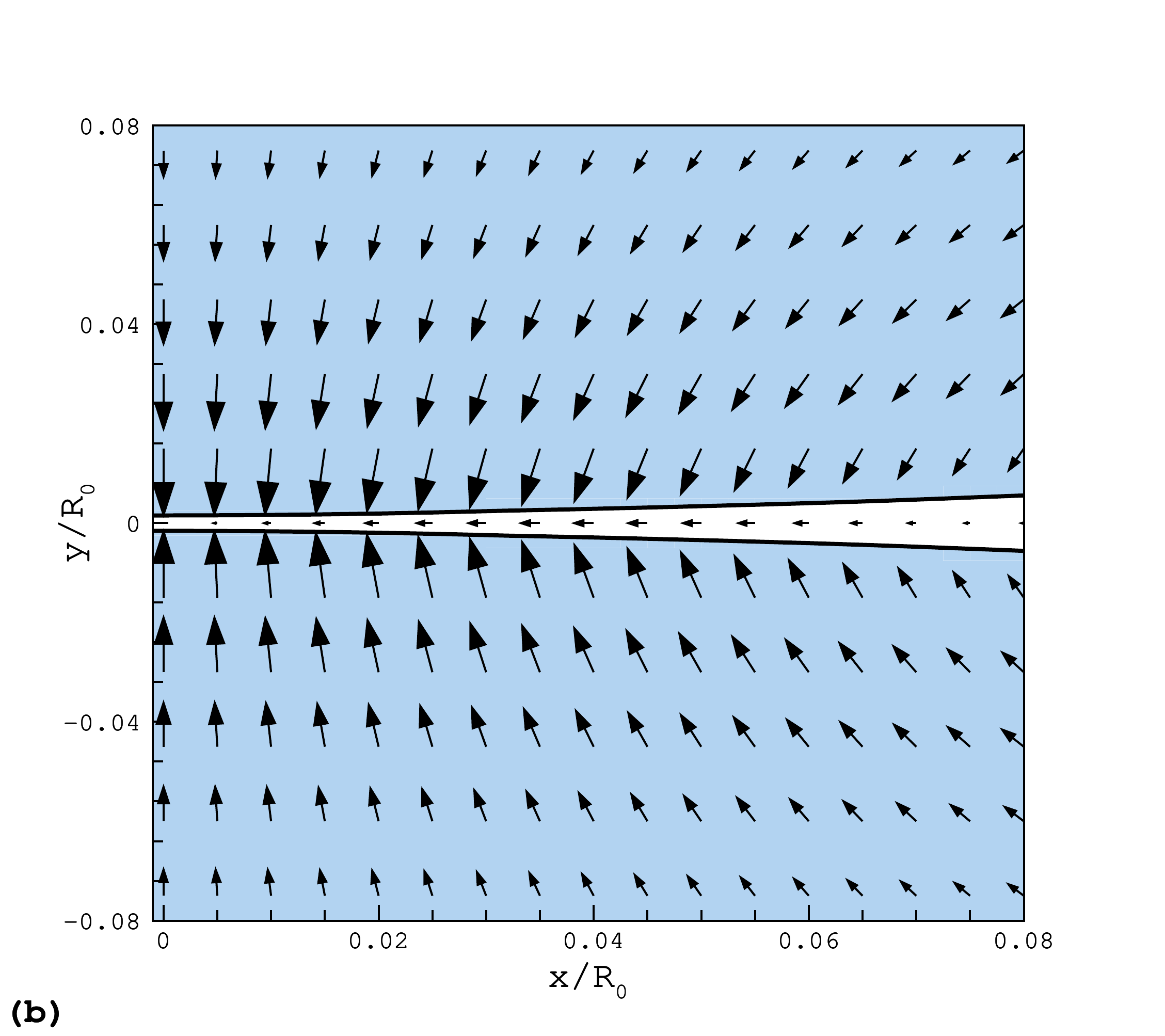}
\end{minipage}

\centering
\begin{minipage}[t]{2cm} %{0.2\linewidth}
\includegraphics[scale=0.28]{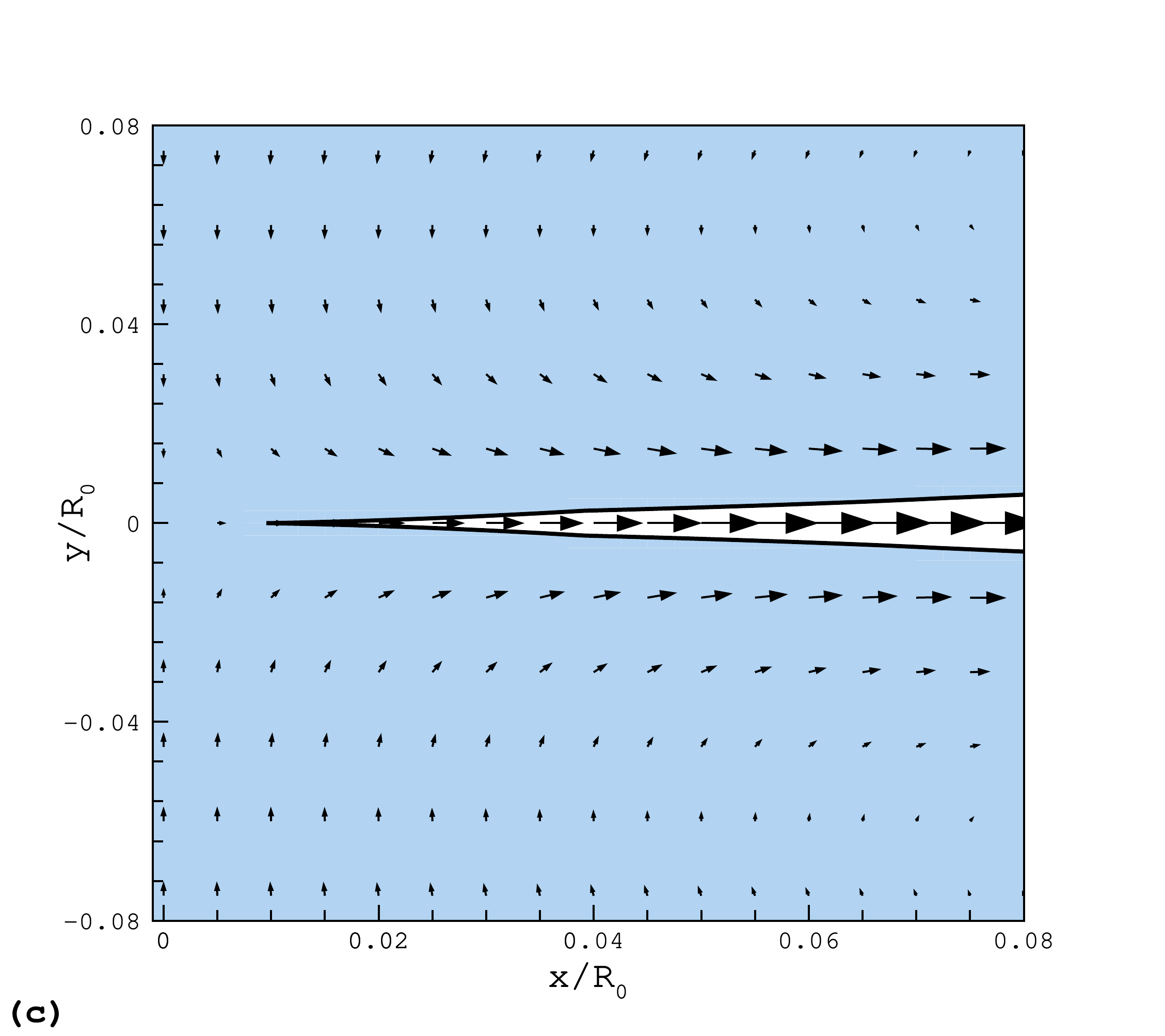}
\end{minipage}
\hspace{5cm}
\begin{minipage}[t]{5.cm} %{0.2\linewidth}
\includegraphics[scale=0.28]{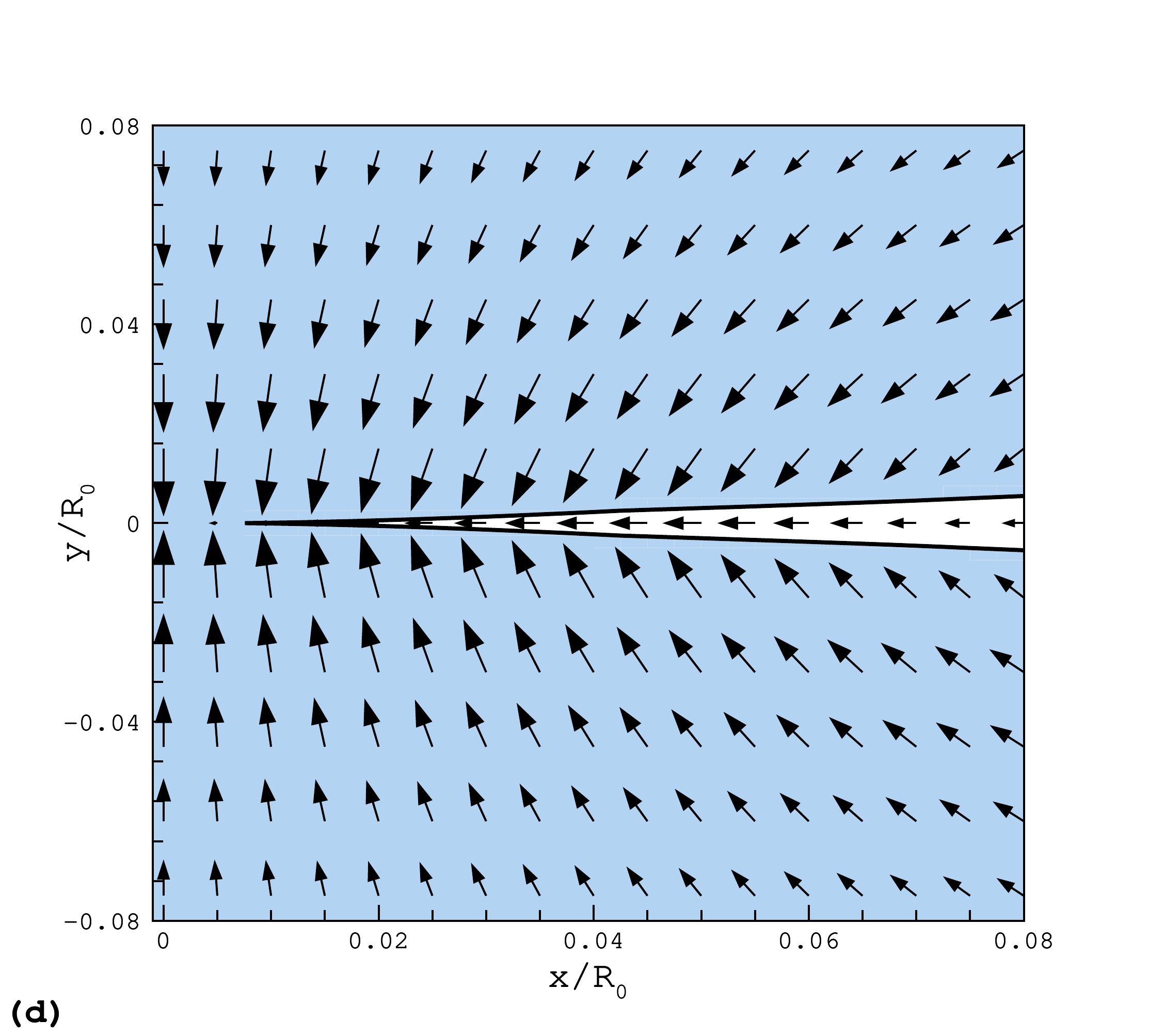}
\end{minipage}

\centering
%\hspace{-2cm}
\begin{minipage}[t]{2.cm} %{0.2\linewidth}
\includegraphics[scale=0.28]{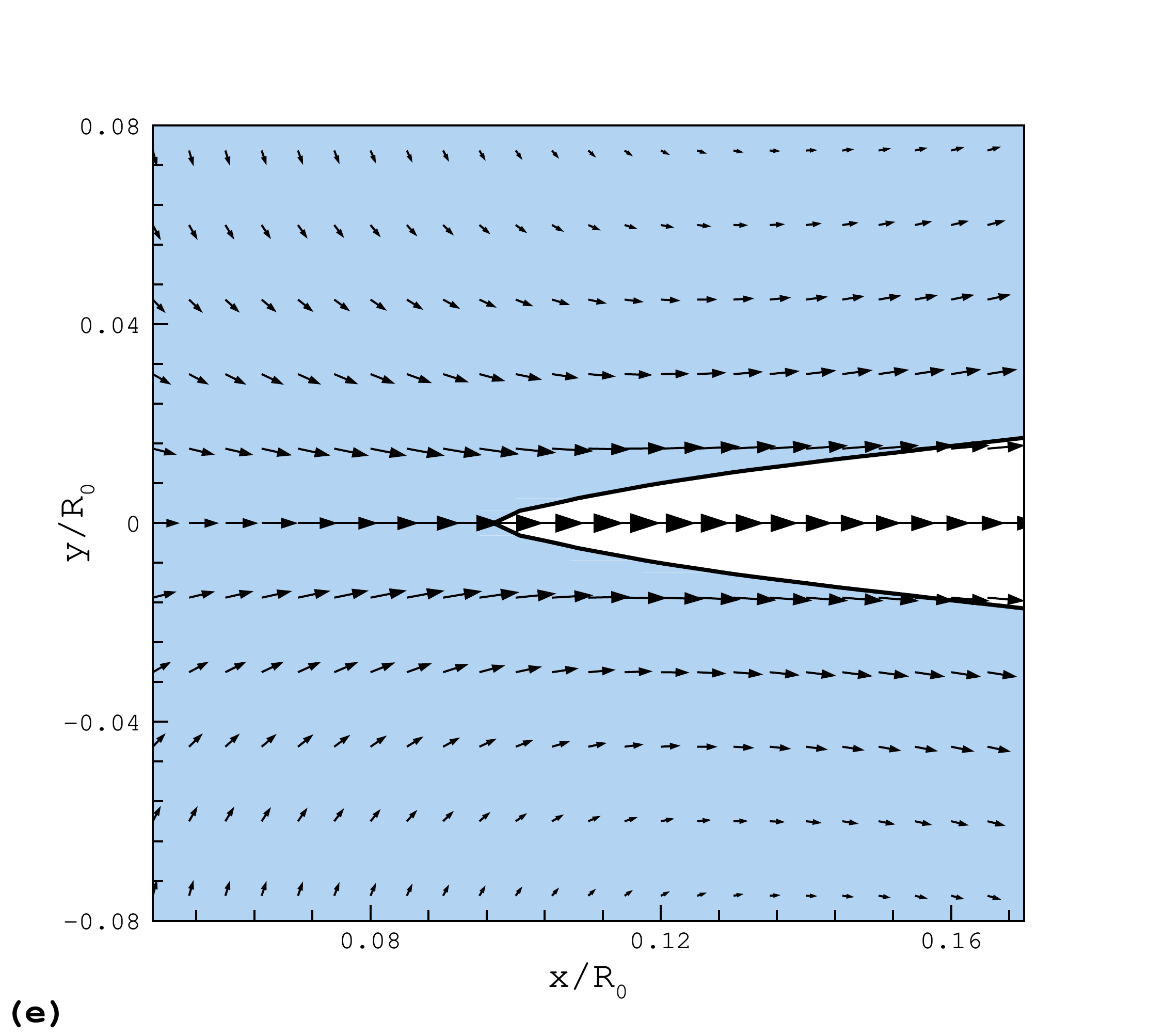}
\end{minipage}
\hspace{5cm}
\begin{minipage}[t]{5.cm} %{0.2\linewidth}
\includegraphics[scale=0.28]{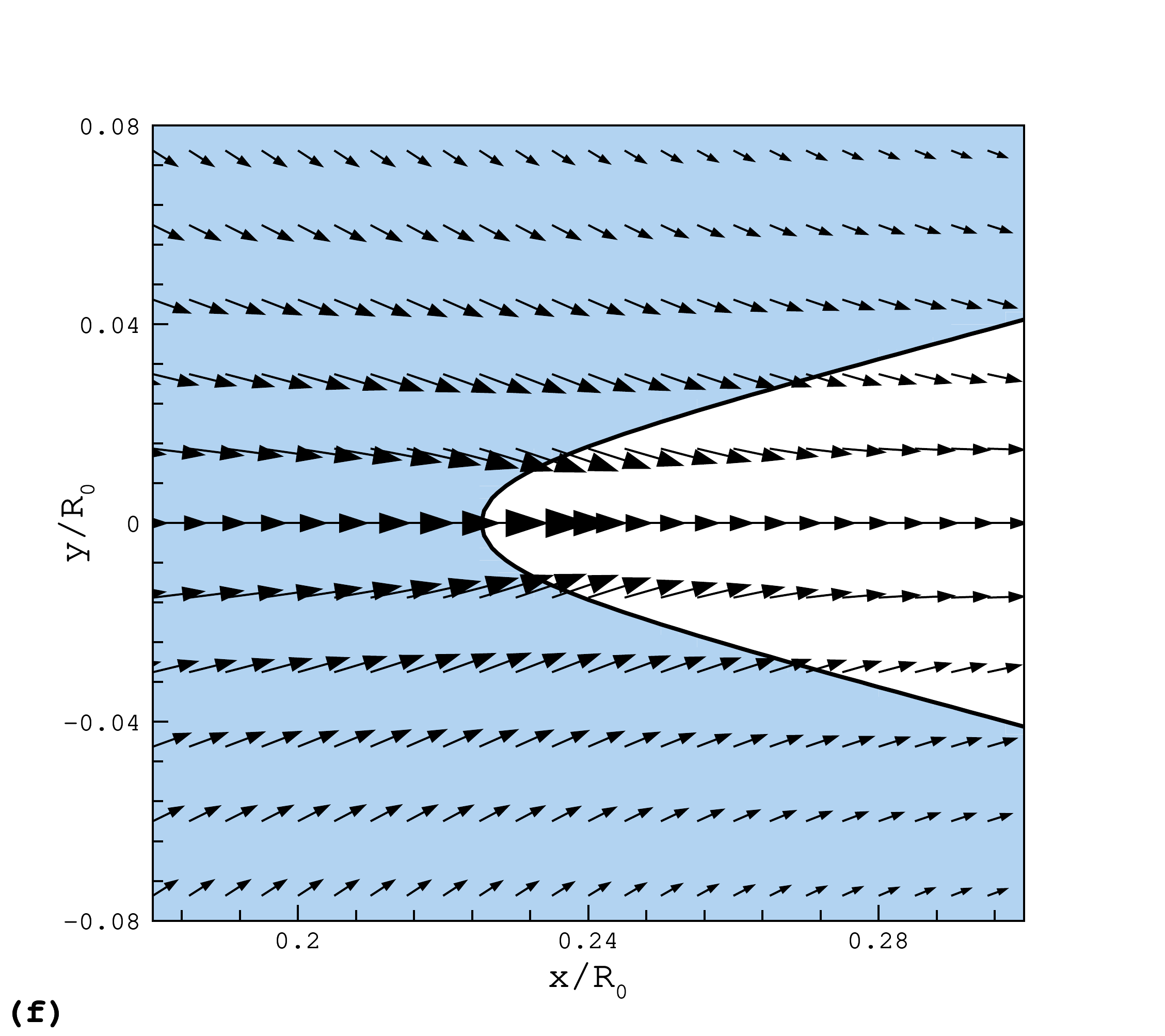}
\end{minipage}
\caption{The velocity field in the region of the liquid bridge before the instant of contact (a,b), at the instant of contact $t/t_{v}=0$ (c,d), and after contact at $t/t_{v}=0.1$ (e,f), resulting from the coalescence simulations in a non-condensable gas (a,c,e), and the coalescence in a saturated vapor phase (b,d,f). The velocity vectors in (a,b), (c,d), (e,f) are plotted with the same scale.}\label{fig:velocity_vectors}
\end{figure}

\begin{figure}
\centering
%\hspace{-2cm}
\begin{minipage}[t]{3.cm} %{0.2\linewidth}
\includegraphics[scale=0.3]{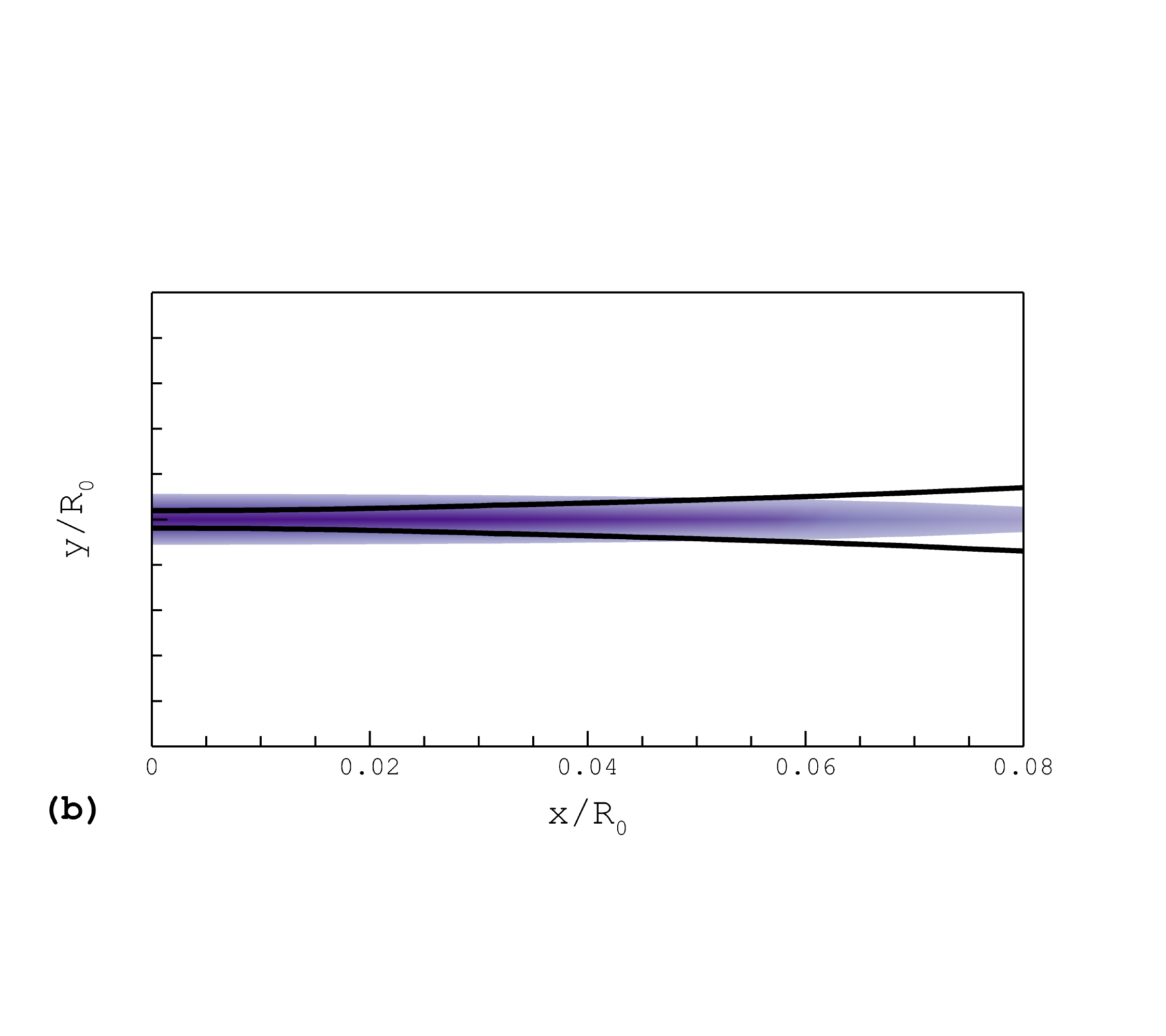}
\end{minipage}
\hspace{5cm}
\begin{minipage}[t]{5.cm} %{0.2\linewidth}
\includegraphics[scale=0.3]{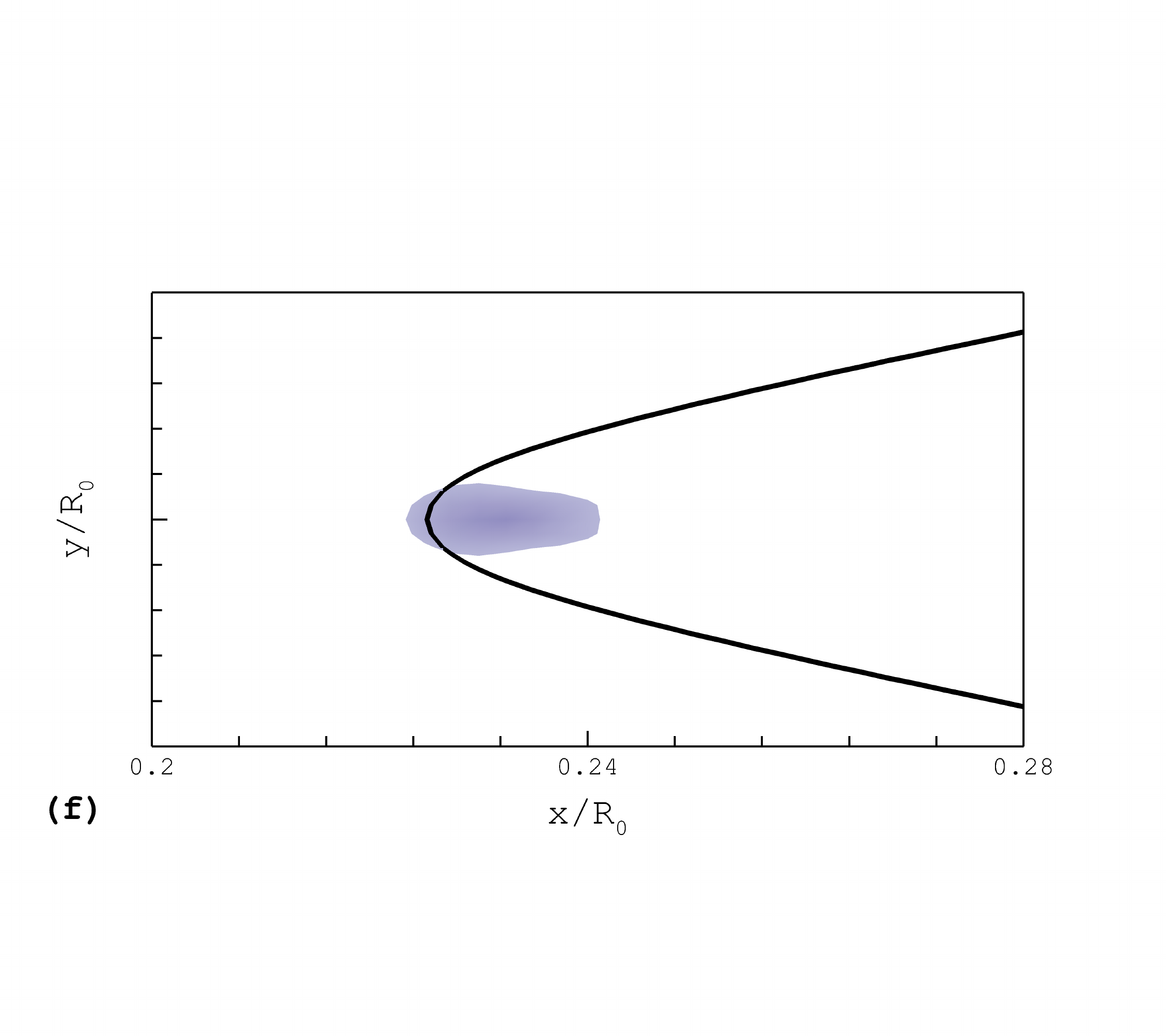}
\end{minipage}
\caption{The color field represents the divergence of the velocity field $\nabla\cdot{\bf u}$ before the instant of contact (b) and after contact at $t/t_{v}=0.1$ (f). The darker color represents the more negative value, \textit{i.e.}, higher condensation rate.}\label{fig:divergence}
\end{figure}

 \begin{figure}
\centering
%\hspace{-2cm}
\begin{minipage}[t]{5.5cm}
\includegraphics[scale=0.27]{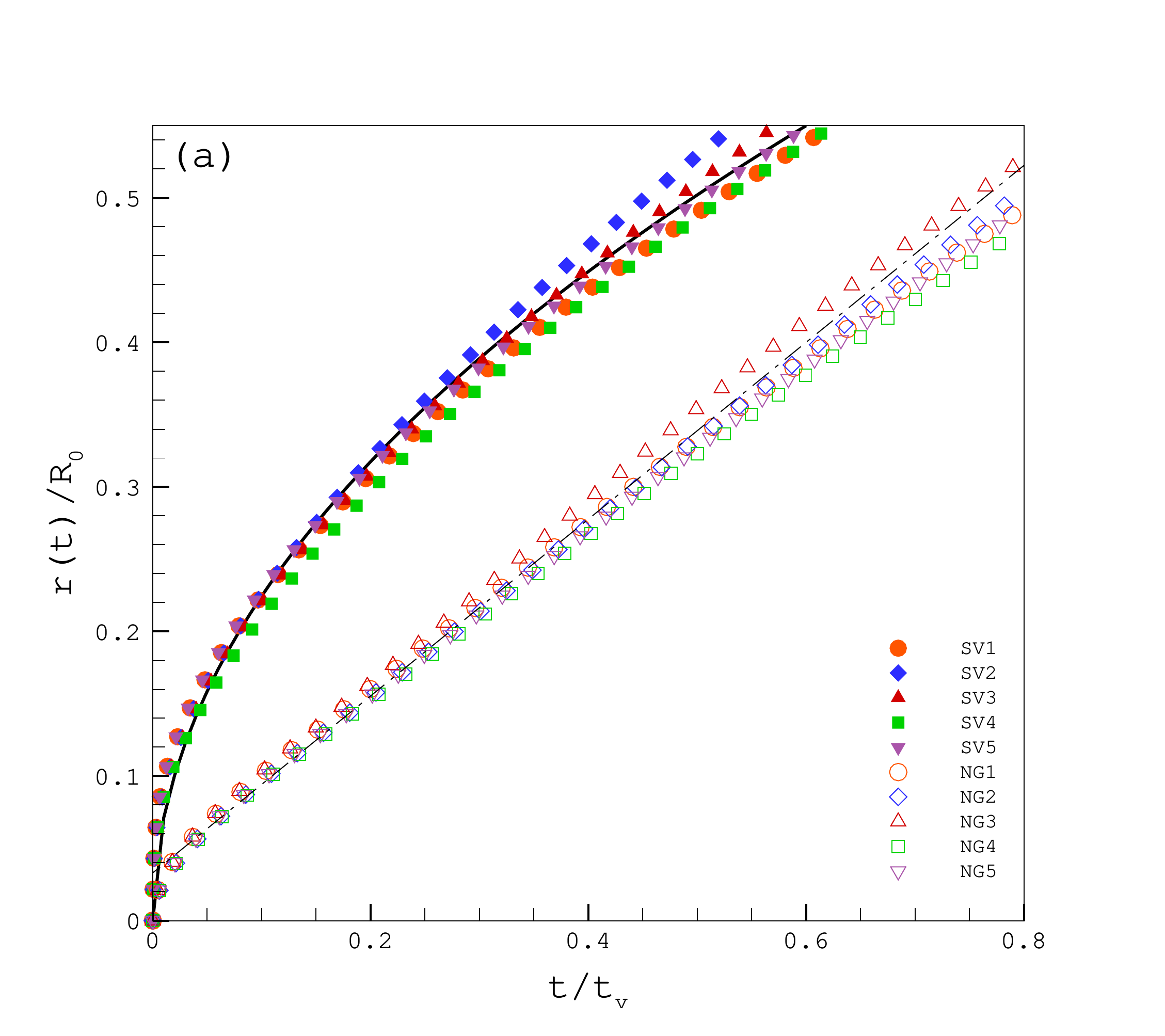}
\end{minipage}
%\hspace{5cm}
\begin{minipage}[t]{5.5cm} %{0.2\linewidth}
\includegraphics[scale=0.27]{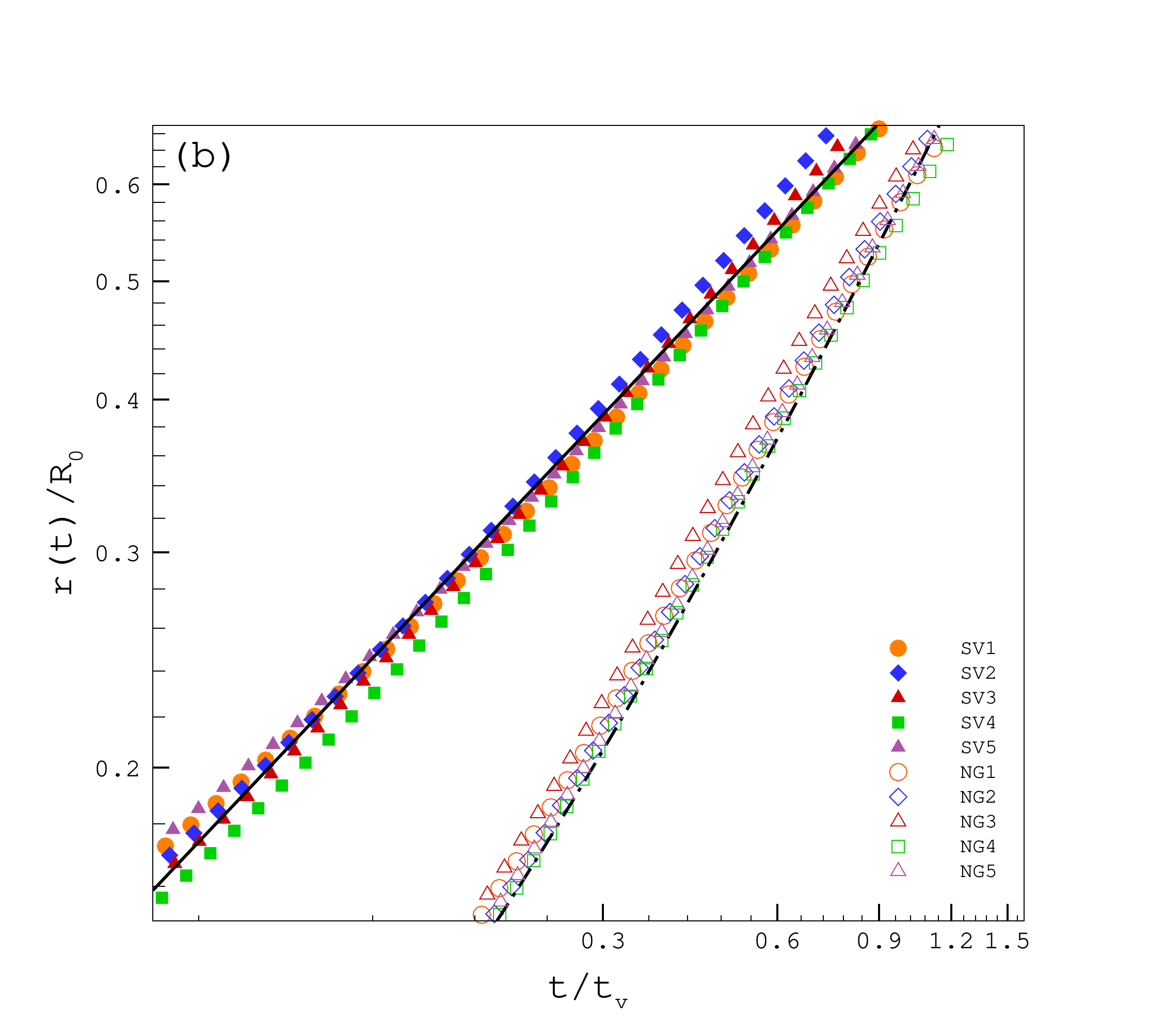}
\end{minipage}
%\hspace{5cm}
\begin{minipage}[t]{5.5cm} %{0.2\linewidth}
\includegraphics[scale=0.27]{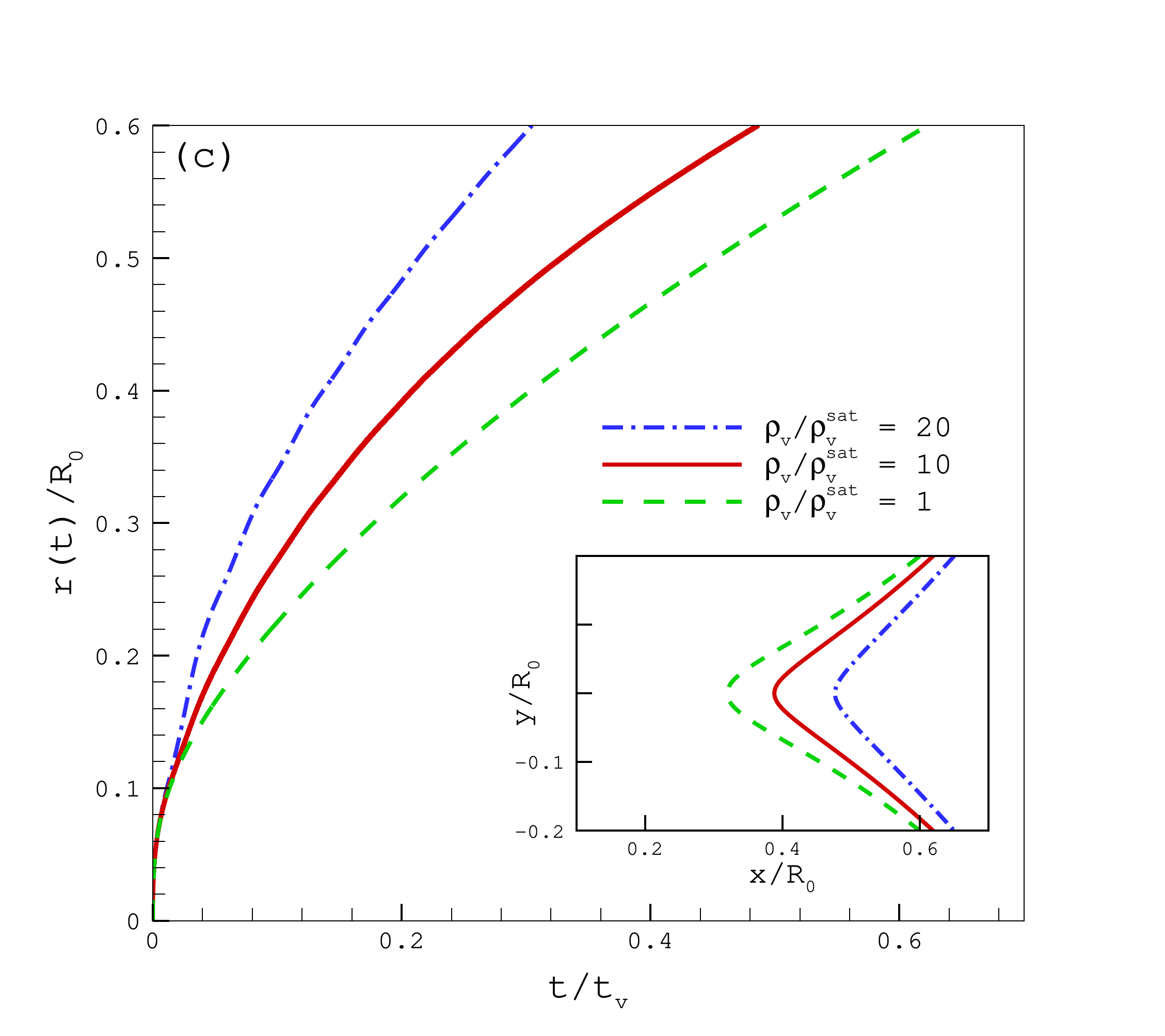}
\end{minipage}
\caption{(a,b) Time evolution of the liquid bridge radius $r(t)$ obtained from the coalescence simulations in a saturated vapor and in a non-condensable gas. The solid symbols represent the data obtained from the coalescence simulations in a saturated vapor (SV), while the open symbols represent the data from the coalescence simulations in a non-condensable gas (NG). In the tests (SV1, NG1), $Oh = 1.2$, $\eta_l/\eta_v =\rho_l/\rho_g = 10$, (SV2, NG2), $Oh = 1.2$, $\eta_l/\eta_v =\rho_l/\rho_g = 100$, (SV3, NG3), $Oh = 0.97$, $\eta_l/\eta_v =\rho_l/\rho_g = 1000$, (SV4, NG4), $Oh = 0.7$, $\eta_l/\eta_v =\rho_l/\rho_g = 100$, and (SV5, NG5), $Oh = 1.8$, $\eta_l/\eta_v =\rho_l/\rho_g = 10$. The solid and the dotted-dashed lines represent power laws $\sim t^{1/2}$ and $\sim t$, respectively. The dotted-dashed line has a slope of~\cite{Aarts_2005} $0.61$. The data resulting from the coalescence in the saturated vapor phase follow $r(t)/R_0\sim (r(t)/t_v)^{1/2}$ (solid line) and the data from coalescence in a non-condensable gas follow $r(t)/R_0\sim t/t_v$ (dotted-dashed line). (c) Effect of the relative saturation ratio $\rho_v/\rho^{sat}_v$ on the time evolution of the liquid bridge. In the three test cases, $Oh = 1.2$ and $\eta_l/\eta_v =100$. Inset: the liquid bridge at $t/t_v = 0.2$ for the studied cases. In (a,c), the data are scaled by the initial drop radius, $R_{0}$, and the viscous time scale, $t_v$. In (b), the data are plotted in a double-logarithmic representation using the same scaling of the axes as in (a,c). }\label{fig:bridge_evoltuion}
\end{figure}

Fig.~\ref{fig:velocity_vectors} compares the velocity field in the vicinity of the liquid bridge (the inset in Fig.~\ref{fig:sketch}) from the coalescence in a non-condensable gas (Figs.~\ref{fig:velocity_vectors}(a),~\ref{fig:velocity_vectors}(c), and~\ref{fig:velocity_vectors}(e)) with that of the coalescence in a saturated vapor phase (Figs.~\ref{fig:velocity_vectors}(b),~\ref{fig:velocity_vectors}(d), and~\ref{fig:velocity_vectors}(f)), before contact (Figs.~\ref{fig:velocity_vectors}(a) and~\ref{fig:velocity_vectors}(b)), at the moment of contact $t/t_{v}=0$ (Figs.~\ref{fig:velocity_vectors}(c) and~\ref{fig:velocity_vectors}(d)), and after contact $t/t_{v}=0.1$ (Figs.~\ref{fig:velocity_vectors}(e) and~\ref{fig:velocity_vectors}(f)). The dimensionless variables for this case are $Oh = 1.2$, $\eta_l/\eta_v =100$, and $\rho_l/\rho_v = 100$. As the two drops are brought together, their interfacial profiles overlap and the intermolecular attraction forces between the two interfaces result in liquid mass flux from the drops toward the interaction zone to initiate the coalescence, as can be observed in Figs.~\ref{fig:velocity_vectors}(a) and~\ref{fig:velocity_vectors}(b). The diffuse interface model incorporates the physics of short-range intermolecular attraction forces~\cite{Baroudi_2014,Pengtao_James_2005}. From Fig.~\ref{fig:velocity_vectors}(a), one can see that, while the droplet surfaces approach each other, the intervening non-condensable gas film proceeds to drain under the influence of the intermolecular attraction forces between the two surfaces, as the velocity vectors in the non-condensable gas are pointing away from the interaction zone. The presence of the non-condensable gas film between the drops slows the liquid motion due to the elevated hydrodynamic pressure in the film. The time required to initiate contact between the two drops in a non-condensable gas is related to the rate of film drainage. On the other hand, the velocity field resulting from the simulation of coalescence in a saturated vapor shows a different trend (Fig.~\ref{fig:velocity_vectors}(b)). The velocity vectors in the saturated vapor phase in the thin film separating the drops are pointing in the opposite direction, toward the interaction zone; see Fig.~\ref{fig:velocity_vectors}(b). For the coalescence in a saturated vapor phase, the intervening vapor film does not thin by drainage of the vapor phase. Instead, we observe mass transfer from the outer vapor phase toward the interaction zone: this  implies condensation. The condensation in our simulations is signaled by the negative divergence of the velocity field $\nabla\cdot{\bf u}<0$. The divergence of the velocity field before the instant of contact is shown in Fig.~\ref{fig:divergence}(b). When the drops are brought together in a condensable vapor near saturation, the intervening vapor film becomes unstable at small separations and phase transition from gas to liquid occurs at separations which define a spinodal for the gas-liquid transition. As the interfacial profiles of the two drops overlap, a gradient in chemical potential (pressure) develops; $\Delta\mu = \mu_{sat} - \mu$ is the positive undersaturation in chemical potential, where $\mu_{sat}$ is the chemical potential at the bulk coexistence. The gradient in the chemical potential induces a diffusional flux $j_{dif}$ from the saturated vapor phase toward the interaction zone, i.e.  condensation takes place. The diffusional flux is written as $j_{dif}=-D\nabla \mu=-D\nabla \left(\partial_\rho E_f\right)$, where $D$ is the diffusion coefficient, and $E_f$ is the free energy of the system. The pressure $p$ is related to the chemical potential by $p = \rho\mu-E_{f}$~\cite{Lee_2006}. In the case of coalescence in a saturated vapor phase, there exist two transport mechanisms that trigger the liquid bridge formation. The first is due to the short range molecular forces that mimic the van der Waals forces in the diffuse interface model, and the second is due to the condensation of the vapor phase at the liquid vapor interface. The onset of coalescence in a saturated vapor phase is observed at $t_s/t_v \approx 1.8\times10^{-3}$, where $t_s$ is the time measured from the beginning of the simulation. However, for the coalescence in a non-condensable gas, phase change is not allowed to take place and the only possible mechanism to connect the drops and lower the system free energy is by draining the gas film separating them under the influence of the interaction forces. The onset of coalescence in a non-condensable gas takes place at $t_s/t_v\approx 33\times10^{-3}$. Thus, the presence of the condensable vapor phase speeds the initiation of the coalescence process.

At the moment of contact, a meniscus forms and a negative Laplace pressure develops in the bridge because of the concave shape of the meniscus. The difference in Laplace pressure between the bulk liquid in the drops and the meniscus drives liquid mass flux from each drop toward the liquid bridge, which expands rapidly to the scale of the drop diameter. From Fig.~\ref{fig:velocity_vectors}(c), we observe that the non-condensable gas escapes radially away from the liquid bridge. However, the velocity field resulting from the coalescence in a saturated vapor (Fig.~\ref{fig:velocity_vectors}(d)) is very different.
%from that seen in Fig.~\ref{fig:velocity_vectors}(c).
In Fig.~\ref{fig:velocity_vectors}(d), the velocity vectors in the saturated vapor phase are directed toward the growing liquid bridge indicating mass transfer across the interface, which represents condensation of the vapor phase at the interface (signaled by $\nabla\cdot{\bf u}<0$). For the coalescence in a saturated vapor, the condensable vapor phase flux is toward the meniscus, because directly over the negatively curved interface of the meniscus, the local vapor pressure of the liquid is determined by the curvature as described by the Kelvin equation: $\ln\left(p_{v}/p_{v}^{sat}\right)=-(\sigma V_{l})/(RT \omega_{k}^{eq})$, where $p_{v}$ is the vapor pressure, $p_{v}^{sat}$ is the saturation vapor pressure, $\omega_{k}^{eq}$ is the equilibrium mean radius of curvature (Kelvin radius), $V_{l}$, $R$, and $T$ are liquid molar volume, ideal gas constant, and temperature, respectively. The difference in vapor pressure will drive diffusional flux of the saturated vapor phase toward the liquid bridge to lower the free energy of the system. This difference is related to the difference between the mean radius of curvature of the bridge meniscus $\omega_{k}$ and the Kelvin radius $\omega_{k}^{eq}$ corresponding to the vapor pressure of the surrounding medium~\cite{Butt_2009}. Thus, when the two drops touch in a saturated vapor phase, the negatively curved meniscus is expected to induce local condensation near the liquid bridge according to the Kelvin equation.
Therefore, when the coalescence takes place in a saturated vapor, as the two drops come into contact, there exist two transport mechanisms that contribute to the liquid bridge growth: the first one is due to the capillary forces and the second is caused by condensation of the vapor phase at the liquid vapor interface. In the case of coalescence in a non-condensable gas, phase change does not occur, and the only mechanism that contributes to the expansion of the liquid bridge is flow resulting from the capillary forces. In the Kelvin equation, $p_{v}^{sat}$, $\sigma$, and $V_{l}$ are all properties of the fluid at equilibrium and are considered constants with respect to the system. Temperature is also considered constant in the Kelvin equation as it is a function of the saturation vapor pressure. Therefore, the variables that govern the condensation rate are the relative vapor pressure $p_v/p_{v}^{sat}$ and the radius of curvature of the meniscus $\omega_{k}$. The condensation of vapor results in the release of latent heat and   increases the temperature difference between the condensed liquid and the surrounding vapor phase. It is unlikely that the heat released by the condensation of vapor could give rise to a large value of $\Delta T$ ~\cite{Kohonen_1999}. Here, we assume large thermal conductivity and isothermal phase change~\cite{Lee2003}.

The bridge radius expands as liquid from the drops moves in, pushing the outer phase to escape the gap (Figs.~\ref{fig:velocity_vectors}(e) and~\ref{fig:velocity_vectors}(f)). From Figs.~\ref{fig:velocity_vectors}(e) and~\ref{fig:velocity_vectors}(f), we notice that the velocity fields resulting from the coalescence in a non-condensable gas (Fig.~\ref{fig:velocity_vectors}(e)) and the coalescence in a saturated vapor (Fig.~\ref{fig:velocity_vectors}(f)) have a similar trend, i.e., liquid from the drops is moving toward the bridge and the outer phase is moving away from the bridge. However, for the coalescence in the saturated vapor phase, there is still condensation (signaled by the negative divergence of the velocity field $\nabla\cdot{\bf u}<0$) at the liquid vapor interface of the highly curved concave meniscus due to the Kelvin effect. The divergence of the velocity field at this stage of the coalescence is shown in Fig.~\ref{fig:divergence}(f). We also observe that the bridge radius has a faster expansion rate in the saturated vapor. Growth of the bridge in the saturated vapor proceeds through the combined effects of capillary advection and condensation.

From the results of Fig.~\ref{fig:velocity_vectors}, two distinct coalescence processes can be identified depending on the vapor pressure of the surrounding medium, in the situation where the two drops approach each other at a vanishingly small velocity and in the absence of thermal fluctuations. First, coalescence can occur at constant liquid volume, when the coalescence process takes place in a medium of negligible vapor pressure (non-condensable gas). In this case, the initiation of coalescence is triggered by mechanical instabilities due to van der Waals forces between the two coalescing drops, and the expansion of the liquid bridge is driven by capillary forces. Second, coalescence can occur with volume change, when the process takes place in a medium of relatively high vapor pressure (condensable vapor phase). In this situation the volume of the two coalescing drops changes as their surfaces approach each other and during the expansion of the meniscus formed between them due to the condensation of the vapor phase. The vapor pressure of the surrounding medium affects the thermodynamic stability of the system~\cite{Chen_2004_prl}.

The validity of the proposed scaling law that governs the evolution of the bridge radius in the viscous regime rests on the assumption that the material transfer to the liquid bridge is solely by capillary forces, without considering any phase change effects. Thus, we expect that the presence of condensation will affect the scaling law governing the expansion of the liquid bridge. In the presence of condensation, the growth rate of the liquid bridge is related to the rate of diffusion of vapor through the liquid phase. If we assume that condensation is diffusion limited, the growth of the bridge radius due to condensation will obey the scaling $r(t)\propto t^{1/2}$. We compare the time evolution of the liquid bridge from the coalescence simulations in a saturated vapor to that in a non-condensable gas in Figs.~\ref{fig:bridge_evoltuion}(a) and~\ref{fig:bridge_evoltuion}(b). Five test cases with varying nondimensional parameters ($Oh$, $\eta_l/\eta_v $, and $\rho_l/\rho_v$) are presented. For all the test cases shown in Figs.~\ref{fig:bridge_evoltuion}(a) and~\ref{fig:bridge_evoltuion}(b), we notice that the time evolution of the liquid bridge resulting from the coalescence in a non-condensable gas follows a linear scaling $r(t)/R_0\sim t/t_v$ as expected for the coalescence of liquid drops in the viscous regime~\cite{Aarts_2005,Paulsen_2011, Burton_2007, Yao_2005} or in the inertially limited viscous regime~\cite{Paulsen_2012}. The linear evolution has a slope varying between $0.58$ and $0.63$ which is in good agreement with the experimental results observed by Aarts \textit{et al.}~\cite{Aarts_2005}. However, all the results obtained from the coalescence in a saturated vapor show a faster time evolution of the liquid bridge and do not follow the linear scaling law $r(t)/R_0\sim t/t_v$. Instead, the bridge radius resulting from coalescence in a condensable vapor approximately follows $r(t)/R_0\sim (t/t_v)^{1/2}$ scaling as expected for diffusion limited condensation. Gross \textit{et al.}~\cite{Gross} recently studied the viscous coalescence of liquid drops in a saturated vapor phase using the non-ideal fluid LBE model~\cite{Lee_2006}. They reported that the liquid bridge radius satisfies $r(t)/R_0\sim (t/t_v)^{1/2}$. However, the expansion speed of the liquid bridge in their results was smaller than what we observe in our simulations. They stated that the reason for deviations of their results from the expected linear scaling $r(t)/R_0\sim t/t_v$ is unknown, and claimed that the effect of condensation was negligible on the time evolution of the liquid bridge. They also proposed a scaling argument to describe their results. As discussed above, condensation drives the outward motion of the liquid bridge in the saturated vapor phase to be initially more rapid than in the non-condensable gas explaining faster time evolution of the liquid bridge. Similarly, condensation results in the deviation of the scaling law from the expected linear scaling.

For the coalescence in a condensable vapor phase, the rate of the bridge expansion is related to the rate of condensation of the vapor phase. Thus, changing the vapor pressure of the outer phase is expected to affect the evolution of the liquid bridge. Fig.~\ref{fig:bridge_evoltuion}(c) shows a comparison between three coalescence simulations in a condensable vapor phase with different relative saturation ratios $p_v/p_{v}^{sat}$. The relative saturation pressure is varied by changing the relative saturation density of the vapor phase $\rho_v/\rho_{v}^{sat}$. As can be seen from Fig.~\ref{fig:bridge_evoltuion}(c), the higher the relative saturation ratio, the faster the evolution of the liquid bridge. Higher vapor pressure enhances the condensation process leading to a faster expansion of the liquid bridge.

In summary, we have numerically investigated the dynamics of viscous coalescence in a saturated vapor phase and in a non-condensable gas. The data obtained from our simulations show that the well-known dynamics of viscous coalescence in vacuum or air change considerably when the coalescence event takes place in a condensable vapor phase, due to condensation during the early stages of the process. Condensation initially drives the coalescence process to proceed faster in time. The presence of condensable vapor in the surrounding medium enhances the coalescence and reduces the time required for merging. Our findings are of both practical and fundamental relevance to the study of the coalescence dynamics in an outer phase of relatively high vapor pressure. Experiments on droplet coalescence in a condensable vapor phase are needed to check the predictions reported here.

This research was supported by the CUNY-University of Chicago MRSEC NSF PREM program DMR-0934206 and at UofC by the NSF MRSEC under DMR-0820054 and NSF Grant DMR-1105145. L.B. gratefully acknowledges financial support from the US Nuclear Regulatory Commission Graduate Fellowship NRC-HQ-12-G-38-0.

% Figures should be put into the text as floats.
% Use the graphics or graphicx packages (distributed with LaTeX2e).
% See the LaTeX Graphics Companion by Michel Goosens, Sebastian Rahtz, and Frank Mittelbach for examples.
%
% Here is an example of the general form of a figure:
% Fill in the caption in the braces of the \caption{} command.
% Put the label that you will use with \ref{} command in the braces of the \label{} command.
%
% \begin{figure}
% \includegraphics{}%
% \caption{\label{}}%
% \end{figure}

% Tables may be be put in the text as floats.
% Here is an example of the general form of a table:
% Fill in the caption in the braces of the \caption{} command. Put the label
% that you will use with \ref{} command in the braces of the \label{} command.
% Insert the column specifiers (l, r, c, d, etc.) in the empty braces of the
% \begin{tabular}{} command.
%
% \begin{table}
% \caption{\label{} }
% \begin{tabular}{}
% \end{tabular}
% \end{table}

% If you have acknowledgments, this puts in the proper section head.
%\begin{acknowledgments}
% Put your acknowledgments here.
%\end{acknowledgments}

% Create the reference section using BibTeX:
%\bibliography{aipauth4-1}

%\begin{figure}
%\centering
%\includegraphics[scale=0.45]{Transfer_mechanism_2_grids_reg_sigma.eps}
%\caption{}\label{fig:touching layout}
%\end{figure}

%\begin{figure}
%\centering
%\includegraphics[scale=0.45]{Transfer_comp_1_point_reg_sigma.eps}
%\caption{}\label{fig:touching layout}
%\end{figure}

\end{document}